\documentclass[twocolumn]{article}

\usepackage[pdftex,breaklinks,colorlinks,
citecolor=blue,
urlcolor=blue]{hyperref}
\usepackage{pstricks}
\usepackage{multimedia}
\usepackage{xcolor}
\usepackage{tikz}
\usetikzlibrary{3d,calc}
\usepackage{float}
\usepackage{cite}
\usepackage{multicol}
\usepackage{subfigure}
\usepackage{tabularx}
\usepackage{booktabs}
\usepackage{graphicx}
\usepackage{amsmath}
\usepackage{amssymb}
\usepackage{latexsym}
\usepackage{mathrsfs}
\usepackage{geometry}
\usepackage{chngcntr}
\usepackage{authblk}
\usepackage{abstract}
\title{Setting the physical scale of dimensional reduction in causal dynamical triangulations}
\author[1,2]{Joshua H. Cooperman}
\author[2,3]{Manuchehr Dorghabekov}
\affil[1]{Department of Physics and Astronomy, Bucknell University, Lewisburg, Pennsylvania, United States}
\affil[2]{Physics Program, Bard College, Annandale-on-Hudson, New York, United States\thanks{Affiliation when we initiated the research reported herein}}
\affil[3]{Applied Mathematics and Informatics Program, American University of Central Asia, Bishkek, Kyrgyz Republic$^{*}$}
\begin{document}

\twocolumn[
\begin{@twocolumnfalse}
\maketitle

\begin{abstract}
Within the causal dynamical triangulations approach to the quantization of gravity, striking evidence has emerged for the dynamical reduction of spacetime dimension on sufficiently small scales. Specifically, the spectral dimension decreases from the topological value of $4$ towards a value near $2$ as the scale being probed decreases. The physical scales over which this dimensional reduction occurs have not previously been ascertained. We present and implement a method to determine these scales in units of either the Planck length or the quantum spacetime geometry's effective de Sitter length. 
We find that dynamical reduction of the spectral dimension occurs over physical scales of the order of $10$ Planck lengths, which, for the numerical simulation considered below, corresponds to the order of $10^{-1}$ de Sitter lengths. 
\end{abstract}

\end{@twocolumnfalse}
]

\saythanks

\emph{Introduction}---Studying the nonperturbative quantization of general relativity afforded by causal dynamical triangulations, Ambj\o rn, Jurkiewicz, and Loll made a striking discovery: the effective dimension of quantum spacetime geometry dynamically reduces to a value near $2$ on sufficiently small scales \cite{JA&JJ&RL7}. This phenomenon---dynamical dimensional reduction---has been independently confirmed within causal dynamical triangulations \cite{RK} and subsequently discovered within other approaches to quantum gravity \cite{SC}. 

Ambj\o rn, Jurkiewicz, and Loll performed numerical measurements of the spectral dimension, a scale-dependent measure of dimensionality as determined by a diffusing random walker. Their measurements yielded the spectral dimension of quantum spacetime geometry as a function of diffusion time, namely the number of steps in the diffusion process. Shorter walks typically probe smaller scales, and longer walks typically probe larger scales, but there is no \emph{a priori} connection between diffusion time and any physical scale. One is thus left pondering the question `Over what physical scales does dynamical reduction of the spectral dimension occur?'. 

After briefly reviewing the formalism of causal dynamical triangulations, the definition of the spectral dimension, and the phenomenology of the former within the latter, we present and implement a method for setting the physical scales of dynamical dimensional reduction. Our method proceeds in two successive steps: we first establish the equivalent of the diffusion time in units of the lattice spacing, and we then establish the equivalent of the lattice spacing in units of either the Planck length or the quantum spacetime geometry's effective de Sitter length. We find that the spectral dimension begins to reduce at a physical scale of $40$ Planck lengths or $0.34$ de Sitter lengths and continues to reduce at least to a physical scale of $10$ Planck lengths or $0.10$ de Sitter lengths. Interestingly, this quantum-gravitational phenomenon occurs on physical scales more than an order of magnitude above the Planck length. 

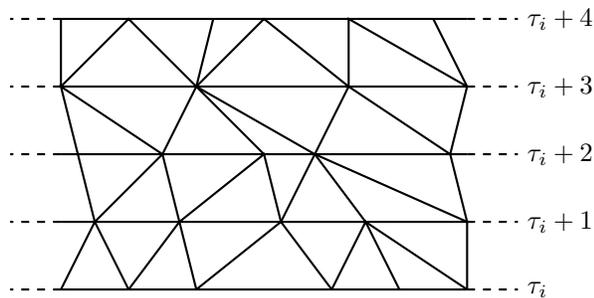
\begin{figure}[ht!]
\centering
\begin{tikzpicture}[scale=0.9]
\draw[thick] (0,0)--(6,0);
\draw[dashed,thick] (0,0)--(-0.75,0);
\draw[dashed,thick] (6,0)--(6.75,0);
\node[right] at (6.75,0) {$\tau_{i}$};
\draw[thick] (0,0)--(0.5,1)--(1,0);
\draw[thick] (1,0)--(1.75,1)--(2,0);
\draw[thick] (2,0)--(3.25,1)--(4,0);
\draw[thick] (4,0)--(4.5,1)--(5,0);
\draw[thick] (4.5,1)--(6,0);
\draw[thick] (6,0)--(6,1);
\draw[thick] (0,1)--(6,1);
\draw[dashed,thick] (0,1)--(-0.75,1);
\draw[dashed,thick] (6,1)--(6.75,1);
\node[right] at (6.75,1) {$\tau_{i}+1$};
\draw[thick] (0.5,1)--(0.25,2);
\draw[thick] (0.5,1)--(1.5,2)--(1.75,1);
\draw[thick] (1.75,1)--(3,2)--(3.25,1);
\draw[thick] (3.25,1)--(3.75,2);
\draw[thick] (4.5,1)--(3.75,2);
\draw[thick] (6,1)--(3.75,2);
\draw[thick] (6,1)--(5.75,2);
\draw[thick] (0,2)--(6,2);
\draw[dashed,thick] (0,2)--(-0.75,2);
\draw[dashed,thick] (6,2)--(6.75,2);
\node[right] at (6.75,2) {$\tau_{i}+2$};
\draw[thick] (0.25,2)--(0,3)--(1.5,2);
\draw[thick] (1.5,2)--(2,3)--(3,2);
\draw[thick] (3.75,2)--(2,3);
\draw[thick] (3.75,2)--(4.25,3);
\draw[thick] (5.75,2)--(4.25,3);
\draw[thick] (5.75,2)--(6,3);
\draw[thick] (0,3)--(6,3);
\draw[dashed,thick] (0,3)--(-0.75,3);
\draw[dashed,thick] (6,3)--(6.75,3);
\node[right] at (6.75,3) {$\tau_{i}+3$};
\draw[thick] (0,3)--(0,4);
\draw[thick] (0,3)--(1,4)--(2,3);
\draw[thick] (2,3)--(2.25,4);
\draw[thick] (2,3)--(3,4)--(4.25,3);
\draw[thick] (4.25,3)--(4.25,4)--(6,3);
\draw[thick] (6,3)--(5.5,4);
\draw[thick] (0,4)--(6,4);
\draw[dashed,thick] (0,4)--(-0.75,4);
\draw[dashed,thick] (6,4)--(6.75,4);
\node[right] at (6.75,4) {$\tau_{i}+4$};
\end{tikzpicture}
\caption{Part of a $2$-dimensional causal triangulation with the discrete time coordinate $\tau$ labeling five consecutive leaves of its distinguished foliation.}
\label{causaltriangulation}
\end{figure}

\emph{Causal dynamical triangulations}---Within a path integral quantization of general relativity, one  formally defines a probability amplitude $\mathscr{A}[\gamma]$ by the equation
\begin{equation}\label{pathintegral}
\mathscr{A}[\gamma]=\int_{\mathbf{g}|_{\partial\mathscr{M}}=\gamma}\mathrm{d}\mu(\mathbf{g})\,e^{iS_{\mathrm{EH}}[\mathbf{g}]/\hbar}:
\end{equation}
integrate over all spacetime metric tensors $\mathbf{g}$, inducing the metric tensor $\gamma$ on the boundary $\partial\mathscr{M}$ of the spacetime manifold $\mathscr{M}$, weighting each by the product of a measure $\mathrm{d}\mu(\mathbf{g})$ and the exponential of $\frac{i}{\hbar}$ times the Einstein-Hilbert action $S_{\mathrm{EH}}[\mathbf{g}]$. Within the causal dynamical triangulations approach to this quantization \cite{JA&AG&JJ&RL3,JA&JJ&RL1,JA&JJ&RL2,JA&RL}, one instead considers a lattice-regularized probability amplitude $\mathcal{A}_{\Sigma}[\Gamma]$ given by the equation
\begin{equation}\label{causalpathsum}
\mathcal{A}_{\Sigma}[\Gamma]=\sum_{\substack{\mathcal{T}_{c}\cong\Sigma\times[0,1] \\ \mathcal{T}_{c}|_{\partial\mathcal{T}_{c}}=\Gamma}}\mu(\mathcal{T}_{c})\,e^{i\mathcal{S}_{\mathrm{R}}[\mathcal{T}_{c}]/\hbar}:
\end{equation} 
sum over all causal triangulations $\mathcal{T}_{c}$ of spacetime topology $\Sigma\times[0,1]$, inducing the triangulation $\Gamma$ on the boundary $\partial\mathcal{T}_{c}$, weighting each by the product of a measure $\mu(\mathcal{T}_{c})$ and the exponential of $\frac{i}{\hbar}$ times the Regge action $\mathcal{S}_{\mathrm{R}}[\mathcal{T}_{c}]$. A causal triangulation $\mathcal{T}_{c}$ is a piecewise-Minkowski simplicial manifold admitting a global foliation by spacelike hypersurfaces all of the chosen topology $\Sigma$. In figure \ref{causaltriangulation} we depict part of a $2$-dimensional causal triangulation. 
One constructs a causal triangulation by appropriately gluing together $N_{D}$ $D$-simplices, each a simplicial piece of $D$-dimensional Minkowski spacetime with spacelike edges of invariant length squared $a^{2}$ and timelike edges of invariant length squared $-\alpha a^{2}$. $a$ is the lattice spacing, and $\alpha$ is a positive constant. As figure \ref{causaltriangulation} shows, these $D$-simplices assemble such that they generate a distinguished spacelike foliation, its leaves labeled by a discrete time coordinate $\tau$. There are $D+1$ types of $D$-simplices; we distinguish these types with an ordered pair $(p,q)$, its entries indicating the numbers of vertices on initial and final adjacent leaves. 

The foliation enables a Wick rotation of a causal triangulation from Lorentzian to Euclidean signature, achieved by analytically continuing $\alpha$ to $-\alpha$ through the lower half complex plane. The probability amplitude \eqref{causalpathsum} transforms accordingly into the partition function
\begin{equation}\label{partitionfunction}
\mathcal{Z}_{\Sigma}[\Gamma]=\sum_{\substack{\mathcal{T}_{c}\cong\Sigma\times[0,1] \\ \mathcal{T}_{c}|_{\partial\mathcal{T}_{c}}=\Gamma}}\mu(\mathcal{T}_{c})\,e^{-\mathcal{S}_{\mathrm{R}}^{(\mathrm{E})}[\mathcal{T}_{c}]/\hbar}
\end{equation} 
in which $\mathcal{S}_{\mathrm{R}}^{(\mathrm{E})}[\mathcal{T}_{c}]$ is the resulting Euclidean Regge action. As in several past studies, we take $\Sigma$ to be the $2$-sphere topology, and we periodically identify the temporal interval $[0,1]$. For these choices
\begin{equation}\label{CDTReggeaction}
\mathcal{S}_{\mathrm{R}}^{(E)}[\mathcal{T}_{c}]=-k_{0}N_{0}+k_{3}N_{3}
\end{equation}
in which $k_{0}$ and $k_{3}$ are specific functions of the bare Newton constant, the bare cosmological constant, $\alpha$, and $a$. We consider the test case of three spacetime dimensions so that the computations required for the analysis presented below are somewhat less intensive. This analysis carries over straightforwardly to the realistic case of four spacetime dimensions, and we fully expect its results to carry over as well since these two cases possess essentially all of the same phenomenology \cite{JA&DNC&JGS&JJ,JA&AG&JJ&RL2,JA&JJ&RL3,JA&JJ&RL7,JA&JJ&RL6,CA&SC&JHC&PH&RKK&PZ,DB&JH,JHC5,JHC&KL&JMM,JHC&JMM,DNC&JJ,RK}. 

We numerically study the partition function \eqref{partitionfunction} for the action \eqref{CDTReggeaction} (at fixed numbers $N_{3}$ of $3$-simplices and $T$ of spacelike leaves) using standard Markov chain Monte Carlo methods. This partition function exhibits two phases of quantum spacetime geometry. We consider exclusively the so-called de Sitter phase, the physical properties of which we discuss below. One ascertains these physical properties by measuring observables $\mathcal{O}_{\mathcal{T}_{c}}$, specifically, their expectation values 
\begin{equation}
\mathbb{E}[\mathcal{O}]=\frac{1}{\mathcal{Z}[\Gamma]}\sum_{\substack{\mathcal{T}_{c}\cong\Sigma\times[0,1] \\ \mathcal{T}_{c}|_{\partial\mathcal{T}_{c}}=\Gamma}}\mu(\mathcal{T}_{c})\,e^{-\mathcal{S}^{(\mathrm{E})}_{\mathrm{cl}}[\mathcal{T}_{c}]/\hbar}\,\mathcal{O}_{\mathcal{T}_{c}}
\end{equation}
in the quantum state defined by this partition function, which we approximate by their averages 
\begin{equation}
\langle\mathcal{O}\rangle=\frac{1}{N(\mathcal{T}_{c})}\sum_{j=1}^{N(\mathcal{T}_{c})}\mathcal{O}_{\mathcal{T}_{c}^{(j)}}
 \end{equation}
over an ensemble of $N(\mathcal{T}_{c})$ causal triangulations generated by our Markov chain Monte Carlo algorithm. 

Ultimately, one aims to learn about the probability amplitudes \eqref{pathintegral} both by taking a continuum limit in which the lattice regularization is removed \emph{via} a nontrivial ultraviolet fixed point and by returning from Euclidean to Lorentzian signature \emph{via} an Osterwalder-Schrader-type theorem.

\emph{Spectral dimension}---The spectral dimension measures the dimensionality of a space as experienced by a random walker diffusing through this space. Taking this space to be a Wick-rotated causal triangulation $\mathcal{T}_{c}$, the spectral dimension is specifically defined as follows \cite{JA&JJ&RL7,JA&JJ&RL6,DB&JH}. 

The integrated discrete diffusion equation
\begin{align}\label{diffusionequation}
\begin{split}
\mathcal{K}_{\mathcal{T}_{c}}(s,s',\sigma)&=(1-\varrho)\mathcal{K}_{\mathcal{T}_{c}}(s,s',\sigma-1)\\ &+\frac{\varrho}{N(\mathscr{N}_{s}(1))}\sum_{s''\in\mathscr{N}_{s}(1)}\mathcal{K}_{\mathcal{T}_{c}}(s'',s',\sigma-1)
\end{split}
\end{align}
governs the random walker's diffusion. The heat kernel element $\mathcal{K}_{\mathcal{T}_{c}}(s,s',\sigma)$ gives the probability of diffusion from $D$-simplex $s$ to $D$-simplex $s'$ (or \emph{vice versa}) in $\sigma$ diffusion time steps. $\mathcal{K}_{\mathcal{T}_{c}}(s,s',\sigma)$ is simply the weighted average of the probability to have diffused from $s$ to $s'$ in $\sigma-1$ steps---the first term on the right hand side of equation \eqref{diffusionequation}---and the probability to diffuse from a $D$-simplex $s''$ in the set $\mathscr{N}_{s}(1)$ of nearest neighbors to $s$ in $\sigma$ steps---the second term on the right hand side of equation \eqref{diffusionequation}. The diffusion constant $\varrho$ characterizes the dwell probability of a step in the diffusion process. By averaging $\mathcal{K}_{\mathcal{T}_{c}}(s,s',\sigma)$ for $s=s'$ over all $N_{s}(\mathcal{T}_{c})$ $D$-simplices in $\mathcal{T}_{c}$, one arrives at the return probability (or heat trace):
\begin{equation}\label{returnprobabilitydef}
\mathcal{P}_{\mathcal{T}_{c}}(\sigma)=\frac{1}{N_{s}(\mathcal{T}_{c})}\sum_{s=1}^{N_{s}(\mathcal{T}_{c})}\mathcal{K}_{\mathcal{T}_{c}}(s,s,\sigma). 
\end{equation}
As its name implies, $\mathcal{P}_{\mathcal{T}_{c}}(\sigma)$---and, subsequently, the spectral dimension---derives from 
random walks that return to their starting $D$-simplices. 

One now defines the spectral dimension $\mathcal{D}_{\mathfrak{s}}^{(\mathcal{T}_{c})}(\sigma)$ as  the power with which $\mathcal{P}_{\mathcal{T}_{c}}(\sigma)$ scales with $\sigma$ multiplied by $-2$:
\begin{equation}\label{specdimdef}
\mathcal{D}_{\mathfrak{s}}^{(\mathcal{T}_{c})}(\sigma)=-2\frac{\mathrm{d}\ln{\mathcal{P}_{\mathcal{T}_{c}}(\sigma)}}{\mathrm{d}\ln{\sigma}}
\end{equation}
for a suitable discretization of the logarithmic derivative. Equation \eqref{specdimdef} provides a measure of a causal triangulation's dimensionality as a function of $\sigma$. 
We approximate the expectation value $\mathbb{E}[\mathcal{D}_{\mathfrak{s}}(\sigma)]$ of $\mathcal{D}_{\mathfrak{s}}^{(\mathcal{T}_{c})}(\sigma)$ 
by the ensemble average $\langle\mathcal{D}_{\mathfrak{s}}(\sigma)\rangle$.
We follow the methods of \cite{JHC} in estimating $\langle\mathcal{D}_{\mathfrak{s}}(\sigma)\rangle$ and its error. 

\begin{figure}[ht!]
\includegraphics[width=\linewidth]{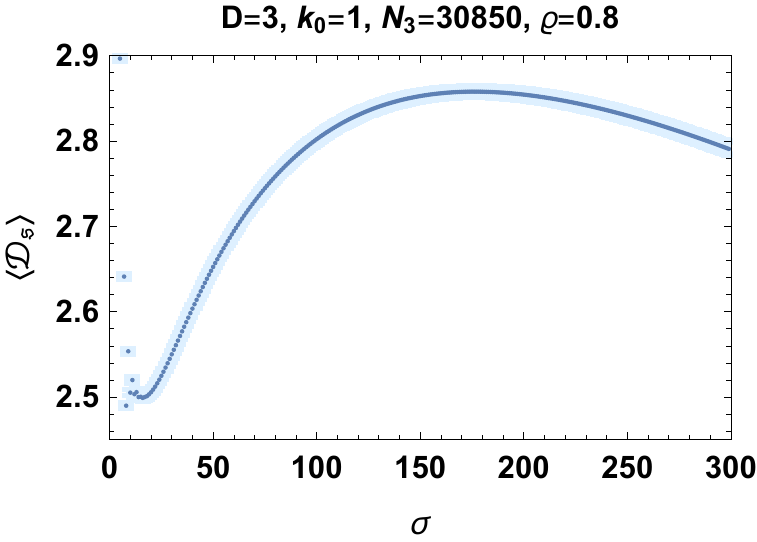}
\caption{The ensemble average spectral dimension $\langle\mathcal{D}_{\mathfrak{s}}\rangle$ as a function of the diffusion time $\sigma$ (in blue). Each point's vertical extent (in light blue) indicates its statistical error .} 
\label{specdim30}
\end{figure}

In figure \ref{specdim30} we display $\langle\mathcal{D}_{\mathfrak{s}}(\sigma)\rangle$ for an ensemble of causal triangulations within the de Sitter phase characterized by $k_{0}=1$ and $N_{3}=30850$ for $\varrho=0.8$. We study this ensemble throughout the paper.\footnote{The analysis that we describe below, particularly its first part, is computationally intensive; accordingly, with the computing resources available to us, we have not yet analyzed ensembles characterized by larger values of $N_{3}$. Cooperman has demonstrated that this ensemble provides physically reliable results for the spectral dimension \cite{JHC5}.} 
The plot in figure \ref{specdim30} displays the characteristic behavior of $\langle\mathcal{D}_{\mathfrak{s}}(\sigma)\rangle$ within this phase. $\langle\mathcal{D}_{\mathfrak{s}}(\sigma)\rangle$ first increases monotonically from a value of approximately $2.5$ to a value of approximately $2.86$ and then decreases monotonically from a value of approximately $2.86$ (eventually) towards a value of $0$. This monotonic rise, followed in reverse, is the phenomenon of dynamical reduction of the spectral dimension; the monotonic fall results from the quantum geometry's large-scale positive curvature \cite{DB&JH}. 
Finite-size effects depress the maximum of $\langle\mathcal{D}_{\mathfrak{s}}(\sigma)\rangle$ below the topological value of $3$ 
\cite{JHC5}. 

\emph{Question}---The diffusion time $\sigma$ is simply the parameter that enumerates the random walker's steps. For smaller values of $\sigma$, the random walker typically probes 
smaller physical scales, and, for larger values of $\sigma$, the random walker typically probes 
larger physical scales. 
The quantitative correspondence between $\sigma$ and the physical scales being probed depends on the space through which the random walker diffuses. 
We propose a method to determine this correspondence for an ensemble of causal triangulations within the de Sitter phase. 
We implement this method to set the physical scales characterizing the phenomenology of the ensemble average spectral dimension $\langle\mathcal{D}_{\mathfrak{s}}(\sigma)\rangle$ within the de Sitter phase. Specifically, 
we determine the interval of physical scales over which dynamical reduction occurs and the physical scale at which $\langle\mathcal{D}_{\mathfrak{s}}(\sigma)\rangle$ coincides with the topological dimension $D$. 


\emph{Methods}---Our method is conceptually straightforward. First we directly determine the average geodesic distance in units of the lattice spacing $a$ traversed by the random walker for walks that return in $\sigma$ diffusion time steps. Then we employ the analysis of \cite{JA&AG&JJ&RL2} to express the lattice spacing $a$ in units of either the Planck length $\ell_{\mathrm{P}}$ or the quantum spacetime geometry's effective de Sitter length $\ell_{\mathrm{dS}}$. 

\begin{figure}[ht!]
\centering
\begin{tikzpicture}[scale=0.9]
\draw[ultra thin,gray] (0,0)--(6,0);
\draw[dashed,ultra thin,gray] (0,0)--(-0.75,0);
\draw[dashed,ultra thin,gray] (6,0)--(6.75,0);
\node[right] at (6.75,0) {$\tau_{i}$};
\draw[ultra thin,gray] (0,0)--(0.5,1)--(1,0);
\draw[ultra thin,gray] (1,0)--(1.75,1)--(2,0);
\draw[ultra thin,gray] (2,0)--(3.25,1)--(4,0);
\draw[ultra thin,gray] (4,0)--(4.5,1)--(5,0);
\draw[ultra thin,gray] (4.5,1)--(6,0);
\draw[ultra thin,gray] (6,0)--(6,1);
\draw[ultra thin,gray] (0,1)--(6,1);
\draw[dashed,ultra thin,gray] (0,1)--(-0.75,1);
\draw[dashed,ultra thin,gray] (6,1)--(6.75,1);
\node[right] at (6.75,1) {$\tau_{i}+1$};
\draw[ultra thin,gray] (0.5,1)--(0.25,2);
\draw[ultra thin,gray] (0.5,1)--(1.5,2)--(1.75,1);
\draw[ultra thin,gray] (1.75,1)--(3,2)--(3.25,1);
\draw[ultra thin,gray] (3.25,1)--(3.75,2);
\draw[ultra thin,gray] (4.5,1)--(3.75,2);
\draw[ultra thin,gray] (6,1)--(3.75,2);
\draw[ultra thin,gray] (6,1)--(5.75,2);
\draw[ultra thin,gray] (0,2)--(6,2);
\draw[dashed,ultra thin,gray] (0,2)--(-0.75,2);
\draw[dashed,ultra thin,gray] (6,2)--(6.75,2);
\node[right] at (6.75,2) {$\tau_{i}+2$};
\draw[ultra thin,gray] (0.25,2)--(0,3)--(1.5,2);
\draw[ultra thin,gray] (1.5,2)--(2,3)--(3,2);
\draw[ultra thin,gray] (3.75,2)--(2,3);
\draw[ultra thin,gray] (3.75,2)--(4.25,3);
\draw[ultra thin,gray] (5.75,2)--(4.25,3);
\draw[ultra thin,gray] (5.75,2)--(6,3);
\draw[ultra thin,gray] (0,3)--(6,3);
\draw[dashed,ultra thin,gray] (0,3)--(-0.75,3);
\draw[dashed,ultra thin,gray] (6,3)--(6.75,3);
\node[right] at (6.75,3) {$\tau_{i}+3$};
\draw[ultra thin,gray] (0,3)--(0,4);
\draw[ultra thin,gray] (0,3)--(1,4)--(2,3);
\draw[ultra thin,gray] (2,3)--(2.25,4);
\draw[ultra thin,gray] (2,3)--(3,4)--(4.25,3);
\draw[ultra thin,gray] (4.25,3)--(4.25,4)--(6,3);
\draw[ultra thin,gray] (6,3)--(5.5,4);
\draw[ultra thin,gray] (0,4)--(6,4);
\draw[dashed,ultra thin,gray] (0,4)--(-0.75,4);
\draw[dashed,ultra thin,gray] (6,4)--(6.75,4);
\node[right] at (6.75,4) {$\tau_{i}+4$};
\draw[dotted,ultra thick] (1.0833,0.667)--(1.5833,0.33)--(2.33,0.667)--(3.0833,0.33)--(3.91667,0.667)--(3.833,1.33)--(4.75,1.33)--(5.1667,1.667)--(4.5833,2.33)--(5.33,2.667)--(4.833,3.33)--(3.833,3.667)--(3.0833,3.33)--(2.41667,3.667)--(1.75,3.667)--(1,3.33)--(1.1667,2.667)--(2.1667,2.33)--(2.0833,1.667)--(1.25,1.33)--(1.0833,0.667);
\node[below] at (1.0833,0.667) {{\footnotesize $s_{0}$}};
\node[above] at (1.5833,0.33) {{\footnotesize $s_{1}$}};
\node[below] at (2.33,0.667) {{\footnotesize $s_{2}$}};
\node[above] at (3.0833,0.33) {{\footnotesize $s_{3}$}};
\node[below] at (3.91667,0.667) {{\footnotesize $s_{4}$}};
\node[above] at (3.833,1.33) {{\footnotesize $s_{5}$}};
\node[below] at (4.75,1.33) {{\footnotesize $s_{6}$}};
\node[left] at (5.1667,1.667) {{\footnotesize $s_{7}$}};
\node[above] at (4.5833,2.33) {{\footnotesize $s_{8}$}};
\node at (5.475,2.8) {{\footnotesize $s_{9}$}};
\node at (4.7,3.2) {{\footnotesize $s_{10}$}};
\node[below] at (3.833,3.667) {{\footnotesize $s_{11}$}};
\node[above] at (3.0833,3.33) {{\footnotesize $s_{12}$}};
\node[below] at (2.41667,3.667) {{\footnotesize $s_{13}$}};
\node at (1.825,3.475) {{\footnotesize $s_{14}$}};
\node at (1.325,3.25) {{\footnotesize $s_{15}$}};
\node[below] at (1.1667,2.667) {{\footnotesize $s_{16}$}};
\node[above] at (2.1667,2.33) {{\footnotesize $s_{17}$}};
\node[below] at (2.0833,1.667) {{\footnotesize $s_{18}$}};
\node at (1.5,1.2) {{\footnotesize $s_{19}$}};
\draw[dotted,thick] (1.0833,0.667)--(0.5,0.33);
\draw[dotted,thick] (2.33,0.667)--(2.667,1.33)--(2.0833,1.667);
\draw[dotted,thick] (1.25,1.33)--(0.75,1.667)--(0.5833,2.33)--(1.1667,2.667);
\draw[dotted,thick] (2.667,1.33)--(3.33,1.667)--(3.833,1.33);
\draw[dotted,thick] (3.91667,0.667)--(4.5,0.33)--(5.1667,0.33)--(5.5,0.667)--(4.75,1.33);
\draw[dotted,thick] (1,3.33)--(0.33,3.667);
\draw[dotted,thick] (3.0833,3.33)--(3.33,2.667)--(2.91667,2.33)--(2.1667,2.33);
\draw[dotted,thick] (2.91667,2.33)--(3.33,1.667);
\draw[dotted,thick] (3.33,2.667)--(4.5833,2.33);
\draw[dotted,thick] (5.25,3.667)--(4.833,3.33);
\draw[dotted,thick] (0.5,0.33)--(-0.1,0.667);
\draw[dotted,thick] (0.5,0.33)--(0.4,-0.33);
\draw[dotted,thick] (1.5833,0.33)--(1.6,-0.33);
\draw[dotted,thick] (3.0833,0.33)--(3,-0.33);
\draw[dotted,thick] (4.5,0.33)--(4.2,-0.33);
\draw[dotted,thick] (5.1667,0.33)--(5.3,-0.33);
\draw[dotted,thick] (5.5,0.667)--(6.3,0.33);
\draw[dotted,thick] (5.1667,1.667)--(6.3,1.667);
\draw[dotted,thick] (5.33,2.667)--(6.2,2.33);
\draw[dotted,thick] (5.25,3.667)--(6.1,3.667);
\draw[dotted,thick] (5.25,3.667)--(5,4.33);
\draw[dotted,thick] (3.833,3.667)--(3.7,4.33);
\draw[dotted,thick] (2.41667,3.667)--(2.5,4.33);
\draw[dotted,thick] (1.75,3.667)--(1.6,4.33);
\draw[dotted,thick] (0.33,3.667)--(0.4,4.33);
\draw[dotted,thick] (0.33,3.667)--(-0.4,3.667);
\draw[dotted,thick] (0.5833,2.33)--(-0.2,2.33);
\draw[dotted,thick] (0.75,1.667)--(0,1.33);
\end{tikzpicture}
\caption{The dual of the part of the $2$-dimensional causal triangulation of figure \ref{causaltriangulation} shown in dotted lines. The thick dotted lines indicate a representative random walk starting from and returning to the $2$-simplex $s_{0}$.}
\label{dualtriangulation}
\end{figure}
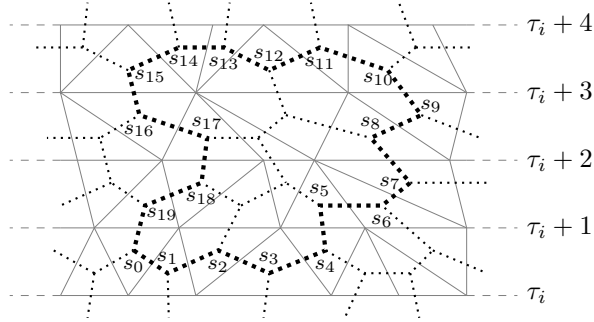

Before presenting our method in detail, we introduce two standard mathematical notions that we use extensively in our method: the dual triangulation and the triangulation geodesic distance. Given a causal triangulation (or, indeed, any triangulation), one constructs its dual in two steps: first place a dual vertex $\tilde{s}$ at the geometric center of each $D$-simplex $s$; then connect $\tilde{s}$ and $\tilde{s}'$ with a dual edge $\tilde{e}_{\tilde{s}\tilde{s}'}$ if and only if $s$ and $s'$ are nearest-neighbor $D$-simplices. In figure \ref{dualtriangulation} we display the dual of the part of the $2$-dimensional causal triangulation depicted in figure \ref{causaltriangulation}. 
As figure \ref{dualtriangulation} shows, a dual triangulation is itself not necessarily a triangulation. One may also conceive of a dual causal triangulation as an abstract mathematical graph. Since the $D$-simplices employed in constructing causal triangulations are not regular, and since every $D$-simplex has $D+1$ nearest-neighbor $D$-simplices, the dual is a weighted $(D+1)$-valent graph. (Of course, one may also conceive of a causal triangulation as an abstract mathematical graph, weighted and polyvalent.) We choose to work with the dual causal triangulation because dual vertices correspond to $D$-simplices, 
rendering diffusion a process along dual edges. 

As a random walker diffuses, hopping from $D$-simplex to $D$-simplex along dual edges, it delineates a path $P$ through the causal triangulation. Let $P\{s,\ldots,s'\}$ be a path from $s$ to $s'$, a string of $D$-simplices. The triangulation distance $d(P\{s,\ldots,s'\})$ of $P\{s,\ldots,s\}$ is the sum of the lengths of the path's dual edges. Denoting by $N_{\tilde{e}_{(p,q)}^{(u,v)}}(P\{s,\ldots,s'\})$ the number of dual edges connecting a $(p,q)$ $D$-simplex and a $(u,v)$ $D$-simplex along $P\{s,\ldots,s'\}$ and by $d(\tilde{e}_{(p,q)}^{(u,v)})$ the length of a dual edge connecting a $(p,q)$ $D$-simplex and a $(u,v)$ $D$-simplex,
\begin{equation}
d(P\{s,\ldots,s'\})=\sum_{\substack{(p,q)\\(u,v)}}N_{\tilde{e}_{(p,q)}^{(u,v)}}(P\{s,\ldots,s'\})\,d(\tilde{e}_{(p,q)}^{(u,v)}).
\end{equation}
If a causal triangulation were regular, then $d(P\{s,\ldots,s'\})$ would simply be the number of dual edges along $P\{s,\ldots,s'\}$ multiplied by the lattice spacing $a$ (multiplied by a number of order $1$). Causal triangulations are not in general regular because $d(\tilde{e}_{(p,q)}^{(u,v)})$ depends on the types of $D$-simplices. For our choice of $\alpha=1$, however, $d(\tilde{e}_{(p,q)}^{(u,v)})=\frac{a}{\sqrt{6}}$ irrespective of the types of $3$-simplices. The triangulation geodesic distance $d_{g}(s,s')$ between $s$ and $s'$ is the minimum of $d(P\{s,\ldots,s'\})$ over the set $\{P\{s,\ldots,s'\}\}$ of paths between $s$ and $s'$:
\begin{equation}
d_{g}(s,s')=\underset{\{P\{s,\ldots,s'\}\}}{\mathrm{min}}d(P\{s,\ldots,s'\}).
\end{equation}
Intuitively, $d_{g}(s,s')$ is the shortest distance (in units of $a$) along dual edges from $s$ to $s'$. 

We now explain the first part of our method in which we establish the lattice distance associated with the diffusion time $\sigma$. A walk that returns to its starting $D$-simplex forms a cycle $C$ in $\widetilde{\mathcal{T}}_{c}$. 
Consider a cycle $C\{s_{0},s_{1},\ldots,s_{\sigma-1}\}$ of $\sigma$ steps starting and ending at $s_{0}$. (Note that we do not include $s_{\sigma}=s_{0}$ in our notation for a cycle.) We associate a distance $\bar{d}_{g}(C\{s_{0},s_{1},\ldots,s_{\sigma-1}\})$ to $C\{s_{0},s_{1},\ldots,s_{\sigma-1}\}$ as follows. We compute $d_{g}(s_{0},s_{k})$ for $k\in\{0,1,\ldots,\sigma-1\}$, 
and we average $d_{g}(s_{0},s_{k})$ over these $k$: 
\begin{equation}
\bar{d}_{g}(C\{s_{0},s_{1},\ldots,s_{\sigma-1}\})=\frac{1}{\sigma}\sum_{k=0}^{\sigma-1}d_{g}(s_{0},s_{k}).
\end{equation}
For the random walk depicted in figure \ref{dualtriangulation}, we list the distances $d_{g}(s_{0},s_{k})$ for $k\in\{0,1,\ldots,19\}$ in table \ref{walktable}. $\bar{d}_{g}(C\{s_{0},s_{1},\ldots,s_{\sigma-1}\})$ is the random walker's average triangulation geodesic distance from its starting $D$-simplex; $\bar{d}_{g}(C\{s_{0},s_{1},\ldots,s_{\sigma-1}\})$ quantifies the typical lattice scale probed by the random walker diffusing along $C\{s_{0},s_{1},\ldots,s_{\sigma-1}\}$. 

\begin{table}
\centering
\begin{tabularx}{7.1cm}{ccccccccccc}
\toprule
$\sigma$ & $d_{g}$ & & $\sigma$ & $d_{g}$ & & $\sigma$ & $d_{g}$ & & $\sigma$ & $d_{g}$\\
\toprule
$0$ & $0$ & & $5$ & $5$ & & $10$ & $8$ & & $15$ & $5$\\
$1$ & $1$ & & $6$ & $6$ & & $11$ & $7$ & & $16$ & $4$\\
$2$ & $2$ & & $7$ & $7$ & & $12$ & $6$ & & $17$ & $3$\\
$3$ & $3$ & & $8$ & $6$ & & $13$ & $7$ & & $18$ & $2$\\
$4$ & $4$ & & $9$ & $7$ & & $14$ & $6$ & & $19$ & $1$\\
\bottomrule
\end{tabularx}
\caption{The triangulation geodesic distances $d_{g}$ of the random walker from its starting simplex $s_{0}$ in units of $a/\sqrt{6}$ as a function of the diffusion time $\sigma$ for the random walk depicted in figure \ref{dualtriangulation}. }
\label{walktable}
\end{table}

As many cycles contribute to the heat kernel element $\mathcal{K}_{\mathcal{T}_{c}}(s_{0},s_{0},\sigma)$, we associate a distance $\bar{d}_{g}(s_{0},\sigma)$ to $s_{0}$ by averaging $\bar{d}_{g}(C\{s_{0},s_{1},\ldots,s_{\sigma-1}\})$ over these $N(C\{s_{0},\sigma\})$ cycles:
\begin{align}
\begin{split}
\bar{d}_{g}(s_{0},\sigma)&=\frac{1}{N(C\{s_{0};\sigma\})}\\ &\times\sum_{j=1}^{N(C\{s_{0};\sigma\})}\bar{d}_{g}(C_{j}\{s_{0},s_{1},\ldots,s_{\sigma-1}\}).
\end{split}
\end{align}
As many $D$-simplices contribute to the return probability $\mathcal{P}_{\mathcal{T}_{c}}(\sigma)$, we associate a distance $\bar{d}_{g}(\sigma)$ to $\mathcal{T}_{c}$ by averaging over all $N_{s}(\mathcal{T}_{c})$ simplices: 
\begin{equation}
\bar{d}_{g}(\sigma)=\frac{1}{N_{s}(\mathcal{T}_{c})}\sum_{s_{0}=1}^{N_{s}(\mathcal{T}_{c})}\bar{d}_{g}(s_{0},\sigma).
\end{equation}
We estimate 
the expectation value $\mathbb{E}[\bar{d}_{g}(\sigma)]$ of $\bar{d}_{g}(\sigma)$ 
by the ensemble average $\langle \bar{d}_{g}(\sigma)\rangle$. 
$\langle \bar{d}_{g}(\sigma)\rangle$ is the distance in units of $a$ that we associate to $\sigma$ for random walks contributing to the ensemble average spectral dimension $\langle\mathcal{D}_{\mathfrak{s}}(\sigma)\rangle$.

The number of cycles, particularly nonsimple cycles, increases tremendously with the diffusion time, so we cannot possibly consider all cycles. To sample cycles efficiently without bias, we explicitly run a computationally reasonable number of random walks. Specifically, for each causal triangulation within an ensemble, we randomly sample of order $10^{2}$ starting $D$-simplices, and, for each sampled starting $D$-simplex, we run of order $10^{2}$ random walks. (Of course, only some of these walks form cycles, and this constitutes the primary inefficiency of our computations.) When estimating the error in our determination of $\langle\bar{d}_{g}(\sigma)\rangle$, we account for the errors stemming from these three levels of sampling. 

\begin{figure}[ht!]
\includegraphics[width=\linewidth]{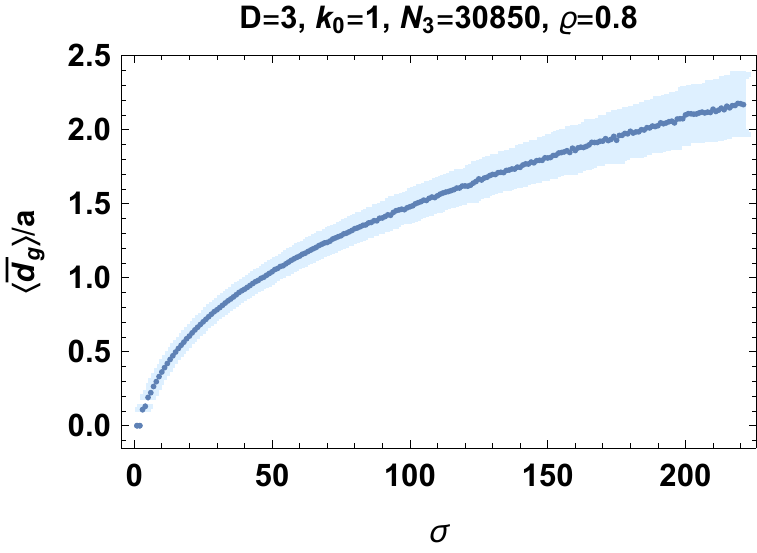}
\caption{The ensemble average geodesic distance $\langle\bar{d}_{g}\rangle$ in units of the lattice spacing $a$ as a function of the diffusion time $\sigma$ (in blue). Each point's vertical extent (in light blue) indicates its statistical error.}
\label{cycledistsigma}
\end{figure}

In figure \ref{cycledistsigma} we display a measurement of $\langle\bar{d}_{g}(\sigma)\rangle$.
By inverting $\langle \bar{d}_{g}(\sigma)\rangle$, 
we determine the scale corresponding to $\sigma$ in units of $a$. The analysis leading to figure \ref{cycledistsigma} constitutes our primary innovation. 

\begin{figure}[ht!]
\includegraphics[width=\linewidth]{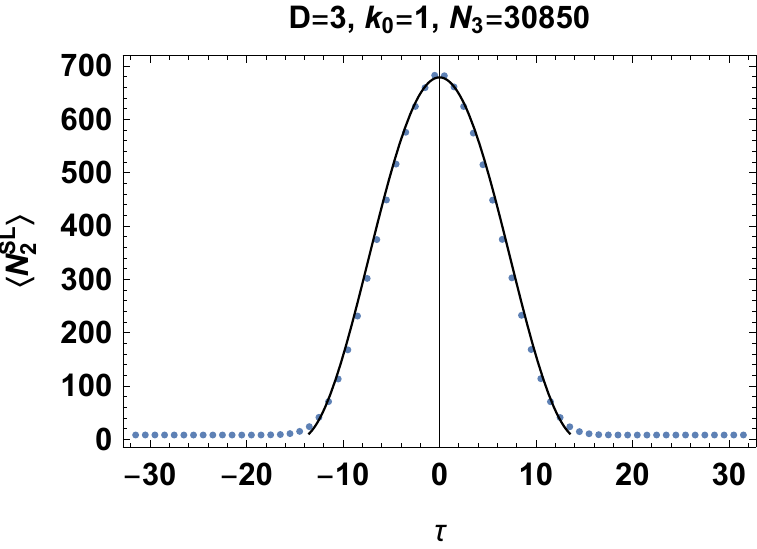}
\caption{The ensemble average number $\langle N_{2}^{\mathrm{SL}}\rangle$ of spacelike $2$-simplices as a function of the discrete time coordinate $\tau$ (in blue) overlain with the best fit discrete analogue $\mathcal{V}_{2}(\tau)$ (in black). Statistical errors are not visible at this plot's scale.}
\label{volumeprofile}
\end{figure}

\begin{figure}[ht!]
\centering
\includegraphics[width=\linewidth]{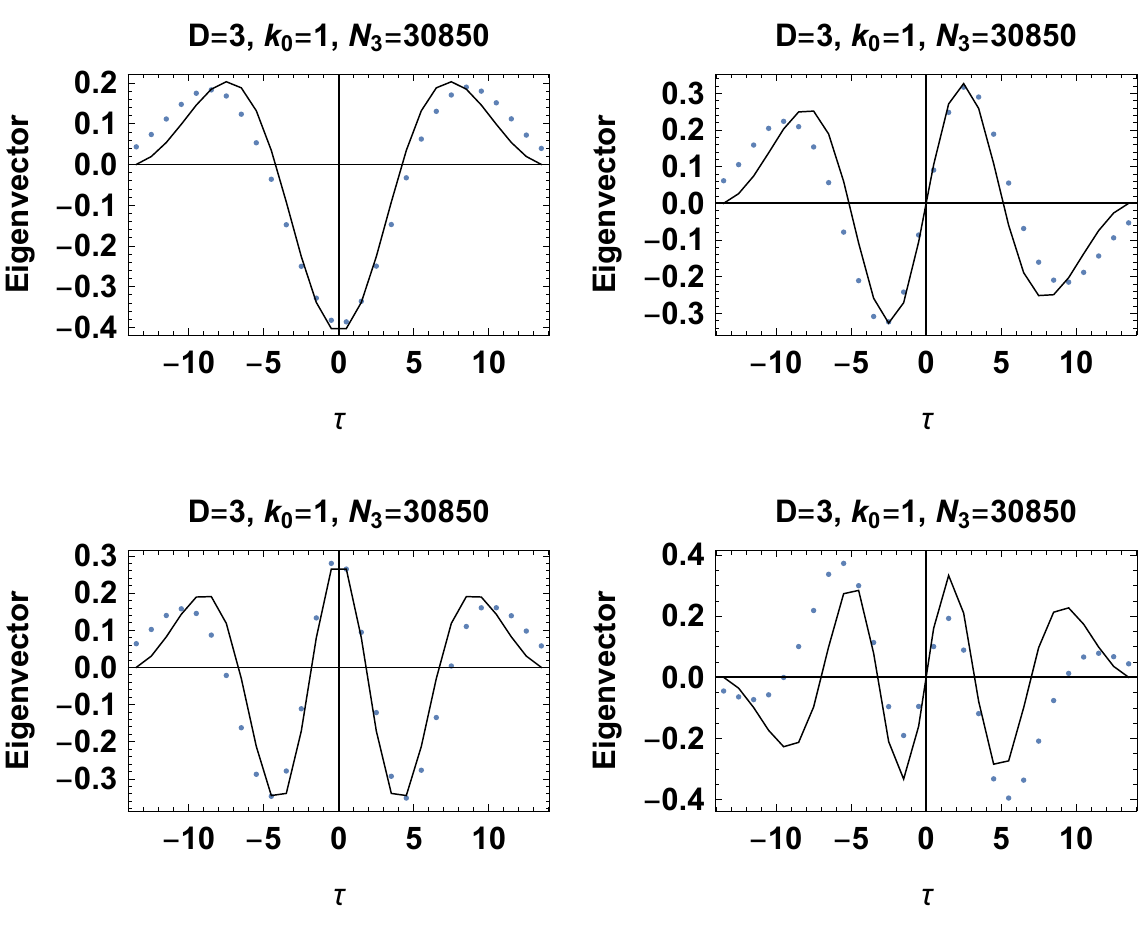}
\caption{The first four eigenvectors (in blue) of $\langle \delta N_{2}^{\mathrm{SL}}(\tau)\,
\delta N_{2}^{\mathrm{SL}}(\tau')\rangle$ overlain with the first four eigenvectors (in black) of the discrete analogue $\langle \delta\mathcal{V}_{2}(\tau)\,\delta\mathcal{V}_{2}(\tau')\rangle$. Statistical errors are not visible at this plot's scale.}
\label{eigenvectors}
\end{figure}

\begin{figure}[ht!]
\centering
\includegraphics[width=\linewidth]{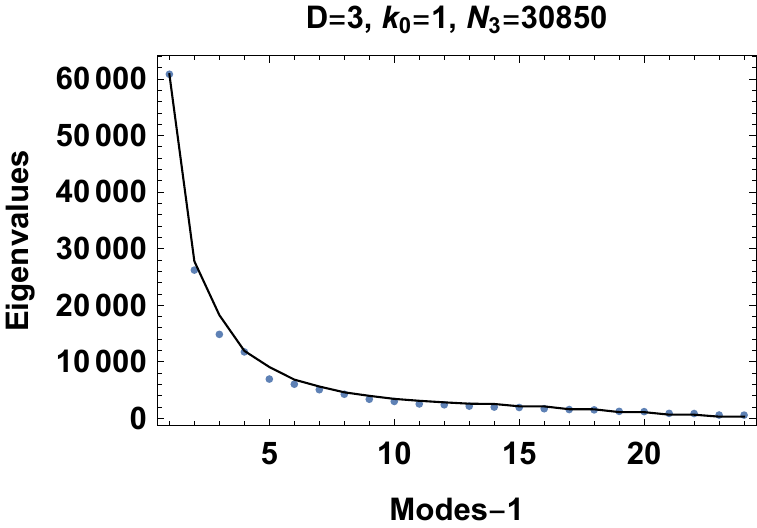}
\caption{The eigenvalues of $\langle \delta N_{2}^{\mathrm{SL}}(\tau)\,
\delta N_{2}^{\mathrm{SL}}(\tau')\rangle$ (in blue) overlain with the eigenvalues (in black) of the discrete analogue $\langle \delta\mathcal{V}_{2}(\tau)\,\delta\mathcal{V}_{2}(\tau')\rangle$. Statistical errors are not visible at this plot's scale.}
\label{eigenvalues}
\end{figure}

We next explain the second part of our method in which we relate the lattice spacing $a$ to two physical length scales---the Planck length $\ell_{\mathrm{P}}$ and the quantum geometry's effective de Sitter length $\ell_{\mathrm{dS}}$---through the analysis first performed for $D=4$ in \cite{JA&AG&JJ&RL2} and subsequently performed for $D=3$ in \cite{JHC&KL&JMM}. These authors analyzed the evolution of the discrete spatial $D$-volume in the distinguished foliation as quantified by the number $N_{D-1}^{\mathrm{SL}}$ of spacelike $(D-1)$-simplices as a function of the discrete time coordinate $\tau$. 
In figure \ref{volumeprofile} we display $\langle N_{2}^{\mathrm{SL}}(\tau)\rangle$ (in blue). Defining the perturbation
\begin{equation}
\delta N_{2}^{\mathrm{SL}}(\tau)=N_{2}^{\mathrm{SL}}(\tau)-\langle N_{2}^{\mathrm{SL}}(\tau)\rangle,
\end{equation} 
we display in figure \ref{eigenvectors} the first four eigenvectors of $\langle \delta N_{2}^{\mathrm{SL}}(\tau)\,\delta N_{2}^{\mathrm{SL}}(\tau')\rangle$ (in blue), and we display in figure \ref{eigenvalues} the eigenvalues of $\langle \delta N_{2}^{\mathrm{SL}}(\tau)\,\delta N_{2}^{\mathrm{SL}}(\tau')\rangle$ (in blue). 

Following \cite{JHC&KL&JMM} in particular, we model $\langle N_{2}^{\mathrm{SL}}(\tau)\rangle$ and $\langle \delta N_{2}^{\mathrm{SL}}(\tau)\,\delta N_{2}^{\mathrm{SL}}(\tau')\rangle$ on the basis of a minisuperspace truncation of the Euclidean Einstein-Hilbert action
\begin{equation}\label{miniaction}
S_{\mathrm{EH}}^{(\mathrm{E})}[V_{2}]=\frac{1}{32\pi G}\int_{t_{\mathrm{i}}}^{t_{\mathrm{f}}}\mathrm{d} t\sqrt{g_{tt}}\left[\frac{\dot{V}_{2}^{2}(t)}{g_{tt}V_{2}(t)}-4\Lambda V_{2}(t)\right]
\end{equation}
(for nonstandard overall sign). $G$ is the renormalized Newton constant, equivalent (for $D=3$) to $\ell_{\mathrm{P}}/\hbar$, and $\Lambda$ is the renormalized cosmological constant. To make direct contact with our measurements of $N_{2}^{\mathrm{SL}}(\tau)$, we express the action \eqref{miniaction} in terms of the spatial $2$-volume $V_{2}$ (as opposed to the scale factor) as a function of the global time coordinate $t$. $\sqrt{g_{tt}}$ is the constant $tt$-component of the metric tensor. 
The extremum of the action \eqref{miniaction} is Euclidean de Sitter space for which
\begin{equation}\label{dSvolprof}
V_{2}^{(\mathrm{EdS})}( t)=4\pi\ell_{\mathrm{dS}}^{2}\cos^{2}{\left(\frac{\sqrt{g_{tt}} t}{\ell_{\mathrm{dS}}}\right)}
\end{equation}
with $ t\in[-\pi\ell_{\mathrm{dS}}/2\sqrt{g_{tt}},+\pi\ell_{\mathrm{dS}}/2\sqrt{g_{tt}}]$ and $\ell_{\mathrm{dS}}=\Lambda^{-1/2}$. $\ell_{\mathrm{dS}}$ is the de Sitter length. Expanding the action \eqref{miniaction} to second order in the perturbation $\delta V_{2}(t)$ about the solution \eqref{dSvolprof},
\begin{align}\label{miniaction2ndorder}
\begin{split}
S_{\mathrm{EH}}^{(\mathrm{E})}[\delta V_{2}]&=S_{\mathrm{EH}}^{(\mathrm{E})}[V_{2}^{(\mathrm{EdS})}]\\ &+\int_{t_{\mathrm{i}}}^{t_{\mathrm{f}}}\mathrm{d}t\int_{t'_{\mathrm{i}}}^{t'_{\mathrm{f}}}\mathrm{d}t' \delta V_{2}(t)\,K(t,t')\,\delta V_{2}(t')\\ &+O\left[\left(\delta V_{2}\right)^{3}\right]
\end{split}
\end{align}
with
\begin{align}
\begin{split}
K(t,t')&=-\frac{\sqrt{g_{tt}}\delta(t-t')}{64\pi^{2}G\ell_{\mathrm{dS}}^{4}}\sec^{2}{\left(\frac{\sqrt{g_{tt}}t}{\ell_{\mathrm{dS}}}\right)}\left[\frac{\ell_{\mathrm{dS}}^{2}}{g_{tt}}\frac{\mathrm{d}^{2}}{\mathrm{d}t^{2}}\right.\\&\left.+\frac{2\ell_{\mathrm{dS}}}{\sqrt{g_{tt}}}\sec{\left(\frac{\sqrt{g_{tt}}t}{\ell_{\mathrm{dS}}}\right)}\tan{\left(\frac{\sqrt{g_{tt}}t}{\ell_{\mathrm{dS}}}\right)}\frac{\mathrm{d}}{\mathrm{d}t}\right.\\&\left.+2\sec^{2}{\left(\frac{\sqrt{g_{tt}}t}{\ell_{\mathrm{dS}}}\right)}\right].
\end{split}
\end{align}
$K(t,t')$ is the van Vleck-Morette determinant \cite{B&D}. A standard calculation of the expectation value $\mathbb{E}[\delta V_{2}(t)\,\delta V_{2}(t')]$ demonstrates that
\begin{equation}\label{propagator}
\mathbb{E}[\delta V_{2}(t)\,\delta V_{2}(t')]=\hbar K^{-1}(t,t').
\end{equation}

This model makes contact with numerical measurements of $N_{2}^{\mathrm{SL}}(\tau)$ through the double scaling limit 
\begin{equation}\label{doublescaling}
V_{3}=\lim_{\substack{N_{3}\rightarrow\infty \\ a\rightarrow 0}}C_{3}N_{3}a^{3}
\end{equation}
for the spacetime $3$-volume $V_{3}$ \cite{JA&AG&JJ&RL2,JA&JJ&RL6,CA&SC&JHC&PH&RKK&PZ,JHC&KL&JMM,JHC&JMM}. In the combination of the thermodynamic ($N_{3}\rightarrow\infty$) and continuum ($a\rightarrow0$) limits, the product $C_{3}N_{3}a^{3}$ approaches a constant, namely $V_{3}$. (For $\alpha=1$, $C_{3}=\frac{1}{6\sqrt{2}}$, the dimensionless discrete spacetime $3$-volume of a $3$-simplex.) Using the double scaling limit \eqref{doublescaling} and the solution \eqref{dSvolprof}, Anderson \emph{et al} \cite{CA&SC&JHC&PH&RKK&PZ}, following \cite{JA&AG&JJ&RL2}, derived the discrete analogue $\mathcal{V}_{2}^{(\mathrm{EdS})}(\tau)$ of the solution \eqref{dSvolprof}:
\begin{equation}
\mathcal{V}_{2}^{(\mathrm{EdS})}(\tau)=\frac{2\langle N_{3}\rangle}{\pi \omega\langle N_{3}\rangle^{1/3}}\cos^{2}{\left(\frac{\tau}{\omega\langle N_{3}\rangle^{1/3}}\right)}
\end{equation}
in which 
\begin{equation}\label{omega}
\omega=\frac{\ell_{\mathrm{dS}}}{\sqrt{g_{tt}}V_{3}^{1/3}}.
\end{equation}
In figure \ref{volumeprofile} we display $\mathcal{V}_{2}^{(\mathrm{EdS})}(\tau)$ (in black) fit to $\langle N_{2}^{\mathrm{SL}}(\tau)\rangle$ (in blue). This first fit determines the value of $\omega$.
Using the double scaling limit \eqref{doublescaling} and the propagator \eqref{propagator}, Cooperman, Lee, and Miller \cite{JHC&KL&JMM}, following \cite{JA&AG&JJ&RL2}, derived the discrete analogue $\langle\delta\mathcal{V}_{2}(\tau)\,\delta\mathcal{V}_{2}(\tau')\rangle$ of the propagator \eqref{propagator}. In figure \ref{eigenvectors} we display the first four eigenvectors of $\langle\delta\mathcal{V}_{2}(\tau)\,\delta\mathcal{V}_{2}(\tau')\rangle$ (in black) fit to the first four eigenvectors of $\langle\delta N_{2}^{\mathrm{SL}}(\tau)\,\delta N_{2}^{\mathrm{SL}}(\tau')\rangle$ (in blue). 
This second fit takes as input the value of $\omega$ determined by the first fit and involves no further fit parameters. In figure \ref{eigenvalues} we display the eigenvalues of $\langle\delta\mathcal{V}_{2}(\tau)\,\delta\mathcal{V}_{2}(\tau')\rangle$ (in black) fit to the eigenvalues of $\langle\delta N_{2}^{\mathrm{SL}}(\tau)\,\delta N_{2}^{\mathrm{SL}}(\tau')\rangle$ (in blue). 
This third fit also takes as input the value of $\omega$ determined by the first fit and also requires the ratio $r$ of the (largest) eigenvalue of $\langle\delta N_{2}^{\mathrm{SL}}(\tau)\,\delta N_{2}^{\mathrm{SL}}(\tau')\rangle$ to the (largest) eigenvalue of $\langle\delta\mathcal{V}_{2}(\tau)\,\delta\mathcal{V}_{2}(\tau')\rangle$. All of these fits improves as $N_{3}$ increases \cite{JHC&JMM}. These fits constitute the primary evidence that the quantum spacetime geometry on sufficiently large scales of the de Sitter phase is that of Euclidean de Sitter space. 

Euclidean de Sitter space has spacetime $3$-volume $V_{3}^{\mathrm{(EdS)}}=2\pi^{2}\ell_{\mathrm{dS}}^{3}$. Substituting $V_{3}^{\mathrm{(EdS)}}$ for $V_{3}$ in the double scaling limit \eqref{doublescaling} (assumed to hold for finite $N_{3}$ and $a$ with negligible corrections), one obtains the relationship
\begin{equation}\label{latticedeSitter}
a=\left(\frac{2\pi^{2}}{C_{3}N_{3}}\right)^{1/3}\ell_{\mathrm{dS}}
\end{equation}
between $a$ and $\ell_{\mathrm{dS}}$. $\mathbb{E}[\delta V_{2}(t)\,\delta V_{2}(t')]$ has eigenvalues proportional to $64\pi^{2}\hbar G\ell_{\mathrm{dS}}^{4}/\sqrt{g_{tt}}$. Relating the eigenvalues of $\mathbb{E}[\delta V_{2}(t)\,\delta V_{2}(t')]$ to the eigenvalues of $\langle \delta N_{2}^{\mathrm{SL}}(\tau)\,\delta N_{2}^{\mathrm{SL}}(\tau')\rangle$ through the double scaling limit \eqref{doublescaling}, and using equations \eqref{omega} and \eqref{latticedeSitter}, one obtains the relationship
\begin{equation}\label{latticePlanck}
a=\frac{32N_{3}^{2/3}}{C_{3}^{1/3}\omega r}\ell_{\mathrm{P}}
\end{equation}
between $a$ and $\ell_{\mathrm{P}}$. 
Having determined $\sigma$ in units of $a$ through our method's first part, we now use equation \eqref{latticedeSitter} or equation \eqref{latticePlanck} to express $a$ in units of $\ell_{\mathrm{dS}}$ or $\ell_{\mathrm{P}}$, finally giving us the ensemble average spectral dimension $\langle\mathcal{D}_{\mathfrak{s}}\rangle$ as a function of a physical scale. 

\emph{Results}---For the ensemble of causal triangulations that we consider, $\omega=0.2978$ and $r=0.0000948$ both with negligible statistical error. Equation \eqref{latticedeSitter} becomes
\begin{equation}
a=0.176\ell_{\mathrm{dS}},
\end{equation}
and equation \eqref{latticePlanck} becomes
\begin{equation}
a=20.46\ell_{\mathrm{P}}.
\end{equation}
Consistent with previous studies, our simulations do not yet probe physical scales below $\ell_{\mathrm{P}}$. 

\begin{figure}[ht!]
\includegraphics[width=\linewidth]{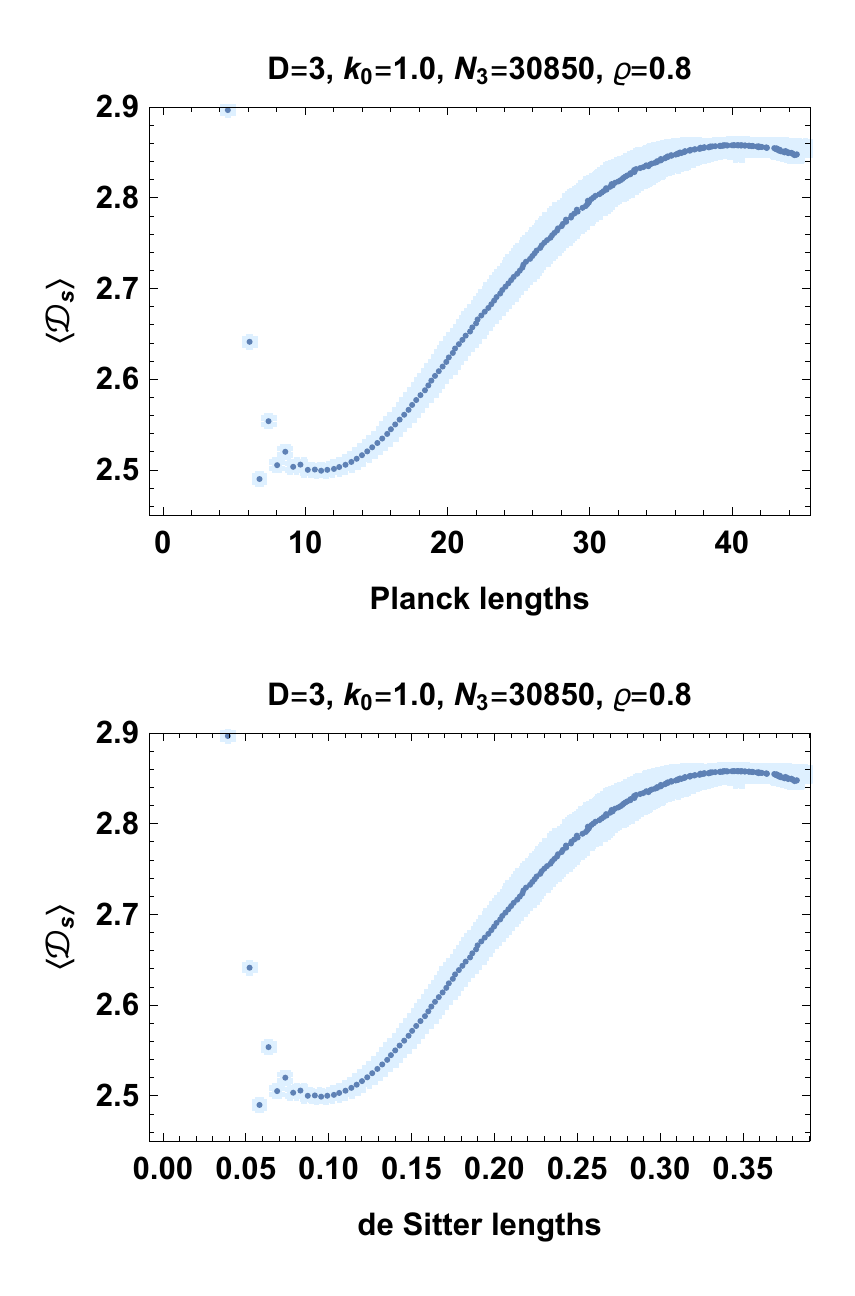}
\caption{The ensemble average spectral dimension $\langle\mathcal{D}_{\mathfrak{s}}\rangle$ as a function of physical scale in units of the Planck length $\ell_{\mathrm{P}}$ (top) and in units of the effective de Sitter length $\ell_{\mathrm{dS}}$ (bottom). Each points' horizontal and vertical extents (in light blue) indicate its statistical error.}
\label{specdimdist}
\end{figure}

In figure \ref{specdimdist} we display the ensemble average spectral dimension $\langle\mathcal{D}_{\mathfrak{s}}\rangle$ as a function of physical scale in units of the Planck length $\ell_{\mathrm{P}}$ and in units of the effective de Sitter length $\ell_{\mathrm{dS}}$.
$\langle\mathcal{D}_{\mathfrak{s}}\rangle$ attains its maximum (depressed below the topological value of $3$ by finite-size effects) at the physical scale of $40\ell_{\mathrm{P}}$ or $0.34\ell_{\mathrm{dS}}$. Dynamical reduction of $\langle\mathcal{D}_{\mathfrak{s}}\rangle$ then extends at least to a physical scale of $10\ell_{\mathrm{P}}$ or $0.10\ell_{\mathrm{dS}}$. 

\emph{Conclusion}---Through a conceptually straightforward but computationally intensive method, we have established the physical scales over which dynamical reduction of the spectral dimension occurs within the de Sitter phase of causal dynamical triangulations. Our analysis demonstrates that this quantum-gravitational phenomenon begins to occurs on physical scales more than an order of magnitude above the Planck length $\ell_{\mathrm{P}}$. Our analysis also demonstrates that the spectral dimension attains the value of the topological dimension $D$ on a physical scale of $40\ell_{\mathrm{P}}$. That the spectral dimension agrees with this value plausibly implies that the quantum spacetime geometry becomes semiclassical on this scale. Such an inference dictates that the quantum spacetime geometry within the de Sitter phase of causal dynamical triangulations is already semiclassical on scales only one order of magnitude above $\ell_{\mathrm{P}}$. Benedetti and Henson's analysis of the spectral dimension indicates that this quantum spacetime geometry is not yet classical on this scale: they found that the ensemble average spectral dimension $\langle\mathcal{D}_{\mathfrak{s}}(\sigma)\rangle$ only begins to match the spectral dimension of Euclidean de Sitter space on a somewhat larger scale \cite{DB&JH}. When combined with our method, Benedetti and Henson's analysis would allow for the determination of the physical scale above which $\langle\mathcal{D}_{\mathfrak{s}}(\sigma)\rangle$ coincides with its classical value and for an independent determination of the quantum geometry's effective de Sitter length $\ell_{\mathrm{dS}}$.

Ambj\o rn, Jurkiewicz, and Loll suggested that $\ell_{\mathrm{P}}$ is the physical scale governing dynamical reduction of the spectral dimension \cite{JA&JJ&RL7}. These authors' suggestion arose from their fit of a phenomenological $3$-parameter function $D_{\mathfrak{s}}(\sigma;\alpha,\beta,\gamma)$ to $\langle\mathcal{D}_{\mathfrak{s}}(\sigma)\rangle$. The  dimensionless parameter $\alpha$ sets $D_{\mathfrak{s}}(\sigma;\alpha,\beta,\gamma)$ to (approximately) $4$ in the limit of large diffusion times; the dimensionless parameter $\beta$ sets $D_{\mathfrak{s}}(\sigma;\alpha,\beta,\gamma)$ to (approximately) $2$ in the limit of small diffusion times; and the dimensionful parameter $\gamma$ determines the rate at which $D_{\mathfrak{s}}(\sigma;\alpha,\beta,\gamma)$ dynamically reduces from $4$ to $2$. Noting that $\gamma$ divides the diffusion time $\sigma$, which itself has dimensions of length squared, they identified $\gamma$ with $\ell_{\mathrm{P}}^{2}$. We interpret Ambj\o rn, Jurkiewicz, and Loll's ensuing discussion as an argument intended to bolster the identification of $\gamma$ with $\ell_{\mathrm{P}}^{2}$. These authors' made two observations. 
First, they estimated the spacetime $4$-volume $V_{4}$ of a causal triangulation in their ensemble as $N_{4}\ell_{\mathrm{P}}^{4}$. We presume that they drew on the double scaling limit 
\begin{equation}\label{doublescaling4}
V_{4}=\lim_{\substack{N_{4}\rightarrow\infty \\ a\rightarrow 0}}C_{4}N_{4}a^{4},
\end{equation}
the equivalent of equation \eqref{doublescaling} for $D=4$. Setting $\ell_{\mathrm{P}}=C_{4}^{1/4}a$ is then an implicit assumption. 
Taking the fourth root of $N_{4}\ell_{\mathrm{P}}^{4}$ yielded approximately $20\ell_{\mathrm{P}}$ for such a causal triangulation's linear size. Second, recalling that $\sigma$ has dimensions of length squared, they estimated a random walker's linear diffusion depth on a causal triangulation in their ensemble as $\sqrt{\sigma}\ell_{\mathrm{P}}$. That one diffusion time step corresponds to a distance $\ell_{\mathrm{P}}$ is essentially the same implicit assumption. 
Considering the diffusion time $\sigma_{\mathrm{max}}$ at which $\langle\mathcal{D}_{\mathfrak{s}}(\sigma)\rangle$ attains a value of $4$ yielded approximately $20\ell_{\mathrm{P}}$ for such a causal triangulation's linear diffusion depth. We presume that they chose to consider $\sigma_{\mathrm{max}}$ on the basis of the previous paragraph's reasoning that the quantum spacetime geometry is plausibly (at least) semiclassical on the scale at which $\langle\mathcal{D}_{\mathfrak{s}}(\sigma)\rangle$ attains a value of $4$. We take Ambj\o rn, Jurkiewicz, and Loll's two observations to imply the argument's unstated conclusion: 
the (approximate) equality of the two estimates constitutes evidence for the validity of the identification of $\ell_{\mathrm{P}}$ with the physical scale governing dynamical reduction of the spectral dimension. 

\begin{figure}[ht!]
\includegraphics[width=\linewidth]{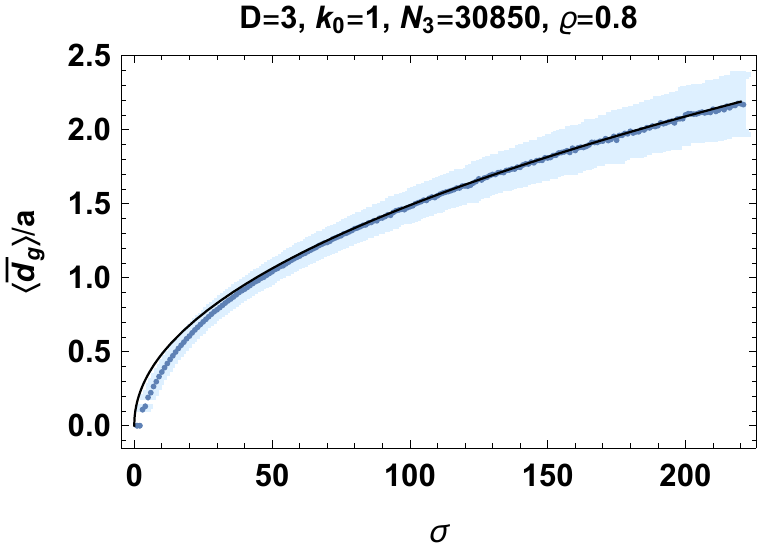}
\caption{The ensemble average geodesic distance $\langle\bar{d}_{g}\rangle$ in units of the lattice spacing $a$ as a function of the diffusion time $\sigma$ (in blue) overlain with the best fit function $\kappa\sigma^{\eta}$ (in black). Each point's vertical extent (in light blue) indicates its statistical error.}
\label{cycledistsigmafit}
\end{figure}

Our above analysis as well as the analyses of Ambj\o rn \emph{et al} \cite{JA&AG&JJ&RL2} and Benedetti and Henson \cite{DB&JH} inform the previous paragraph's argument. The implicit assumption---that $\ell_{\mathrm{P}}=C_{4}^{1/4}a$---yields $\ell_{\mathrm{P}}\approx \frac{1}{2}a$ for typical values of $C_{4}$. In combination with the double scaling limit \eqref{doublescaling4} and the spacetime 4-volume of Euclidean de Sitter space, the estimate of $V_{4}$ yields $\ell_{\mathrm{dS}}\approx 3a$. Ambj\o rn \emph{et al}'s more detailed analysis corroborates these estimates \cite{JA&AG&JJ&RL2}. 
Ambj\o rn, Jurkiewicz, and Loll's estimate of the linear diffusion depth then dictates that $\langle\mathcal{D}_{\mathfrak{s}}(\sigma)\rangle$ attains a value of $4$ on a scale of approximately $3\ell_{\mathrm{dS}}$. This value is an order of magnitude greater than the same scale's value, $0.34\ell_{\mathrm{dS}}$, within our simulations. Moreover, Benedetti and Henson's analysis suggests that $\sigma$ reaches the scale $\ell_{\mathrm{dS}}$ well beyond $\sigma_{\mathrm{max}}$, the value of $\sigma$ at which $\langle\mathcal{D}_{\mathfrak{s}}(\sigma)\rangle$ attains the value $D$ \cite{DB&JH}. One might therefore suspect that estimating the linear diffusion depth as $\sqrt{\sigma}$---the scaling for Euclidean space---is simply too naive; however, our measurement of the ensemble average geodesic distance $\langle\bar{d}_{g}(\sigma)\rangle$ justifies this estimate. Fitting the function $\kappa \sigma^{\eta}$ to $\langle\bar{d}_{g}(\sigma)\rangle$ yields $\kappa=0.157\pm0.001$ and $\eta=0.488\pm0.002$ for these two parameters. In figure \ref{cycledistsigmafit} we display $\kappa\sigma^{\eta}$ (in black) fit to $\langle\bar{d}_{g}(\sigma)\rangle$ (in blue). 
The plot in figure \ref{cycledistsigmafit} shows that $\langle\bar{d}_{g}(\sigma)\rangle$ increases with $\sigma$ very nearly as $\sqrt{\sigma}$ except for sufficiently small $\sigma$. We have thus substantiated Ambj\o rn, Jurkiewicz, and Loll's estimates.

While $N_{4}^{1/4}\ell_{\mathrm{P}}$ and $\sqrt{\sigma_{\mathrm{max}}}\ell_{\mathrm{P}}$ agree for the ensemble of $4$-dimensional causal triangulations that Ambj\o rn, Jurkiewicz, and Loll considered, 
$N_{3}^{1/3}\ell_{\mathrm{P}}$ and $\sqrt{\sigma_{\mathrm{max}}}\ell_{\mathrm{P}}$ disagree by an order of magnitude for the ensemble of $3$-dimensional causal triangulations that we consider. The argument based on the approximate equality of $N_{D}^{1/D}\ell_{\mathrm{P}}$ and $\sqrt{\sigma_{\mathrm{max}}}\ell_{\mathrm{P}}$ breaks down for $D=3$, and we now doubt that this argument holds generally for $D=4$. This breakdown notwithstanding, we can lend new support to Ambj\o rn, Jurkiewicz, and Loll's suggestion that $\ell_{\mathrm{P}}$ governs dynamical reduction of the spectral dimension. Above we have unveiled the following picture: within simulations studied so far for $D=3$, dynamical reduction of $\langle\mathcal{D}_{\mathfrak{s}}(\sigma)\rangle$ occurs over scales of order $10\ell_{\mathrm{P}}$ or $10^{-1}\ell_{\mathrm{dS}}$, and, within simulations studied so far for $D=4$, dynamical reduction of $\langle\mathcal{D}_{\mathfrak{s}}(\sigma)\rangle$ occurs over scales of order $10\ell_{\mathrm{P}}$ or $\ell_{\mathrm{dS}}$. 
The physical scale characterizing dynamical reduction of $\langle\mathcal{D}_{\mathfrak{s}}(\sigma)\rangle$ is independent of $D$ when expressed in units of $\ell_{\mathrm{P}}$, which suggests that $\ell_{\mathrm{P}}$ sets the scale of this quantum-gravitational phenomenon.

Cooperman first advocated that measurements of $\langle\mathcal{D}_{\mathfrak{s}}(\sigma)\rangle$ could form the basis for a renormalization group analysis of causal dynamical triangulations, and he proposed a method for performing such an analysis \cite{JHC}. Subsequently, Ambj\o rn \emph{et al} attempted to track relative changes in the lattice spacing across the de Sitter phase with measurements of $\langle\mathcal{D}_{\mathfrak{s}}(\sigma)\rangle$ \cite{JA&DNC&JGS&JJ}. These authors employed a different method, which Cooperman criticized \cite{JHC4}. Our above analysis, when combined with Cooperman's scaling analysis of the spectral dimension \cite{JHC5}, should allow for the realization of Cooperman's original proposal. We hope that our analysis thereby aids the search for a continuum limit of causal dynamical triangulations effected by a nontrivial ultraviolet fixed point of the renormalization group.

\emph{Acknowledgements}---We thank Christian Anderson, Jonah Miller, and especially Rajesh Kommu for allowing us to employ parts of their codes. 
We also thank Steve Carlip, Hal Haggard, and Jonah Miller for useful discussions. We acknowledge the hospitality and support of the Physics Program of Bard College where we completed much of this research.

\end{document}